\documentclass[aps,prd,reprint,groupedaddress,preprintnumbers,showpacs]{revtex4-1}

\usepackage{graphics} 
\usepackage{amsbsy}

\begin{document}

\preprint{JLAB-PHY-17-2605}

\title{Radiative effects in deep virtual Compton scattering}


\author{Igor Akushevich}
\email[]{igor.akushevich@duke.edu}
\altaffiliation{Jefferson Lab., Newport News, VA 23606, USA}
\affiliation{Physics Department, Duke University, Durham, NC 27708, USA}
\author{Alexander Ilyichev}
\email[]{ily@hep.by}
\affiliation{Institute for Nuclear Problems, Belarusian State University, Minsk, 220030, Belarus}

\date{\today}

\begin{abstract}
 Radiative corrections to the cross sections of photon electroproduction and the single spin asymmetries induced 
 by the interference between the Bethe Heitler and deep virtual Compton scattering amplitudes 
 are calculated within the leading log approximation. The deep virtual Compton scattering amplitude is 
presented in the Belitsky, M\"uller, and Kirchner approximation for the polarized initial particles. The Fortran code for estimation of the radiative effects in a given kinematic point and Monte Carlo generator for simulation of one or two photons are developed. 
Numerical results are performed for beam-spin asymmetries in kinematical conditions of current experiments in the Jefferson Laboratory.
\end{abstract}


\maketitle
\section{\label{Intro}Introduction}

\begin{figure}[t]\centering
\scalebox{0.32}{\includegraphics{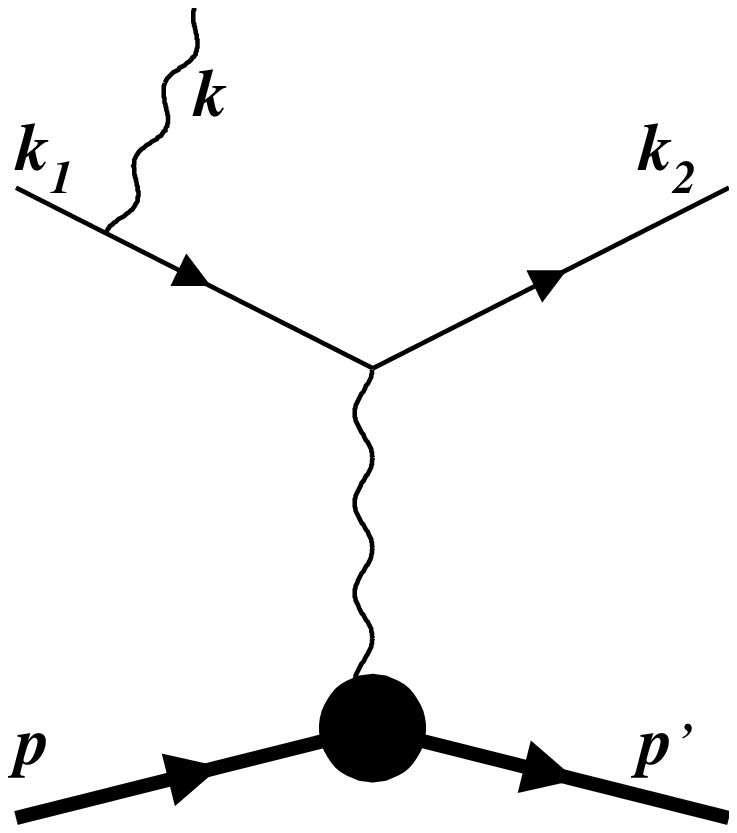}}
\hspace{0.4cm}
\scalebox{0.32}{\includegraphics{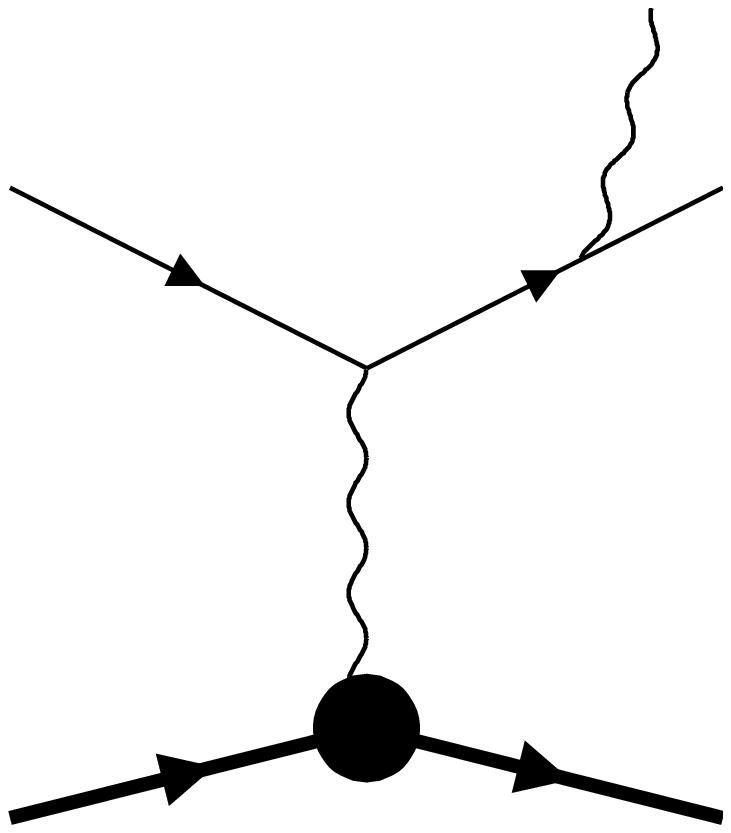}}
\\
{\bf a) \hspace{2cm} b)}
\\
\scalebox{0.32}{\includegraphics{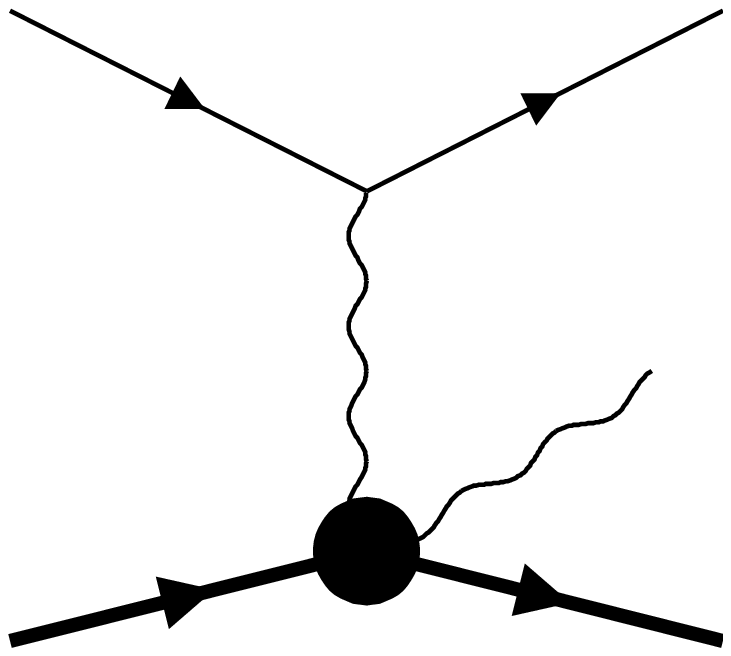}}
\\
{\bf c)}
\caption{\label{DVCSgraphs}Feynman graphs of BH (a and b) and DVCS process (c).}
 \end{figure}

The process of deep virtual Compton scattering
(DVCS) is considered to provide useful information
for extraction of properties of the generalized parton
distributions. Experimentally DVCS is investigated through the measurements of the cross section and asymmetries in the processes of the photon electroproduction with both unpolarized and polarized electron beam and proton target.  Three Feynman graphs presented in Figure \ref{DVCSgraphs} 
contribute
to the cross section of the photon electroproduction. The graphs a) and b) represent the amplitude of the Bethe-Heitler (BH) process and the graph c) describes the DVCS amplitude. The latter gives the access to the properties of the generalized parton distributions, therefore it is of specific interest. During the last decade the process was intensively investigated both theoretically \cite{BelitskyMuller2009PRD,BeMu2010PRD} and experimentally \cite{Girodetal2008PRL}. The cross section of the photon electroproduction is dominated by the BH process, i.e., by the sum of two BH amplitudes (graphs a) and b)) squared.  Therefore, to get an access to DVCS process the researcher has to find an asymmetry vanishing for pure BH process and for which the main contribution would involve the DVCS 
amplitude. 
The well-known example of an appropriate asymmetry is the single beam spin asymmetry. 

The QED radiative correction (RC) is one serious source of systematical uncertainties and therefore must be known with  any predetermined accuracy. Available calculations of QED radiative effects in \cite{Vanderhaeghen2000,ByKuTo2008PRC,AkushevichIlyichev2012,AISh2014}
are focused on RC to BH process or have certain limitations and cannot cover needs of modern requirements on the photon electroproduction. In this paper we present the radiative correction calculations to the cross section of BH and DVCS processes in leading approximation. In the approximation the only leading term containing $L=\log(Q^2/m^2)$ ($m$ is the electron mass)
 is 
kept.   
Since the structure of the dependence of the RC cross section
on the electron mass is 
$
\sigma_{RC}=A\log({Q^2}/{m^2})+B+O(m^2/Q^2),
$
where $A$ and $B$ do not depend on the electron mass, the used approximation allows to keep the major part of the RC, e.g., for experiments with transferring momentum squared of one GeV squared ($Q^2 \sim$ 1GeV$^2$), $\log\bigl({Q^2}/{m^2}\bigr) \sim 15$.

Calculation of RC for BH and DVCS requires knowledge of hadronic structure for these processes. Although the BH cross section can be described in the model-independent way using the nucleon form factors, the description of DVCS amplitude requires assumptions on a model for hadronic subprocesses. A reasonable approach leading to analytic and transparent results is the BMK approximation \cite{BKM2002} in which  quantitative
estimates for the azimuthal and spin asymmetries can be obtained relying on a simple ansatz for the generalized parton distributions. The theoretical consideration went beyond the leading twist and involved the complete analysis in the twist-three approximation. In practice the results obtained in the BMK approximation are considered valid for 
small squared of the transfered momentum between initial and final proton $t$, $-t\leq$0.5 GeV$^2$.

The paper is organized as follows.  The lowest order contribution to the cross section for the photon
electroproduction induced by the interference of BH and DVCS amplitudes
is represented in Section \ref{Born}. Specific attention is paid on explicit representation of the  cross section including polarization part of the cross section, 
mass corrections, as well as for angular structure of the BH cross section. RC calculation is performed in Section \ref{RC}. First we calculate the matrix element squared and trace all sources of occurrence of the electron mass dependence. We keep the contributions of BH amplitude squared and the interference between the BH and DVCS amplitudes, and drop pure DVCS contribution. Second, we represent the phase space of two final photons, introduce so-called shifted kinematics, and calculate integrals over additional photon phase space. Third, we add the contribution of loops and calculate the lowest order RC to the cross section of BH and DVCS processes. Fourth, we generalize the result for the RC to the BH/DVCS cross section to represent the higher order  corrections.  
Section \ref{codes} presents the codes for numerical calculation of RC in a kinematical point and the Monte Carlo generator allowing for simulating the events with one or two radiated photon(s). Section \ref{SectNumeric} provides  numeric estimates of the radiative effects in current experiments in JLab focusing on the RC 
to the cross section in wide kinematic region of modern experiments at JLab.  Finally in Section \ref{SectDiscussion} we summarize the results obtained in this paper.

\section{\label{Born}The cross sections of the interference of BH and DVCS amplitudes}

The process of interest is 
\begin{equation}\label{BHprocess}
e(k_1,\xi)+p(p,\eta)\longrightarrow e'(k_2)+p'(p')+\gamma(k),
\end{equation}
where $k_1^2=k_2^2=m^2$, $p^2=p'^2=M^2$, $k^2=0$, $\xi$ and $\eta$ are polarization vectors of the initial lepton and proton, 
$m$ and $M$ are their masses, respectively.  
The process (\ref{BHprocess}) 
is traditionally described 
by the 
five kinematical variables: 
$S=2k_1p$, $Q^2=-(k_1-k_2)^2$, $x=Q^2/(2p(k_1-k_2))$, $t=(p-p')^2$, and $\phi$ (the angle between 
$({\bf k_1},{\bf k_2})$ and $({\bf q},{\bf p'})$  planes, $q=k_1-k_2$). Four latter variables involve the momenta of final particles, therefore, the cross section of interest is $\sigma\equiv d\sigma_0/dQ^2dxdtd\phi$. Because of azimuthal symmetry the integration over lepton angle $\phi_e$ (i.e., the angle between 
$({\bf k_1},{\bf k_2})$ and $({\bf q},{\bf k_2})$  planes) has been completed in this cross section. The symmetry can be violated in the case of transversal target polarization, therefore the five dimensional cross section is considered $\sigma\equiv d\sigma_0/dQ^2dxdtd\phi(d\phi_e/2\pi)$ in this case. Explicitly, 
the cross section of the photon electroproduction involving the contributions of BH and DVCS amplitudes is  
\begin{eqnarray}\label{dGamma}
d\sigma_{1\gamma}&=&\frac 1{2S} {\cal M}_{1\gamma}^2 d\Gamma_0
\nonumber\\
&=&\frac 1{2S} \bigl({\cal M}_1^l+{\cal M}_2^l+{\cal M}^h\bigr)^2 d\Gamma_0,
\end{eqnarray}
where ${\cal M}_1^l$ and ${\cal M}_2^l$ corresponds to diagrams presented in Figure \ref{DVCSgraphs}a and b, and  ${\cal M}^h$ described hadronic emission contribution presented in Figure \ref{DVCSgraphs}c.

 Phase space for the BH cross section is parameterized as 
\begin{eqnarray}
d\Gamma_0&=&\frac 1{(2\pi)^5}
\frac{d^3k_2}{2E_2}
\frac{d^3p'}{2p'_0}
\frac{d^3k}{2\omega}
\delta^4(k_1+p-k_2-p'-k)
\nonumber \\&=&
\frac{Q^2dQ^2dxdtd\phi}{(4\pi)^4x^2 S \sqrt{\lambda _Y}}
\end{eqnarray}
with $\lambda _Y=S_x^2+4 M^2Q^2$ and $S_x=S-X=Q^2/x$. Kinematical limits on $t$ are defined as
\begin{eqnarray}
t_{2,1}=-\frac{1}{2W^2}\bigl((S_x-Q^2)(S_x\pm\sqrt{\lambda_Y})+2M^2Q^2\bigr),
\end{eqnarray}
where $W^2=S_x-Q^2+M^2$.

All variables are ultimately expressed in terms of the five kinematical variables: $S$, $t$, $Q^2$, $x$ and $\phi$, e.g.,
\begin{eqnarray}
w_0&=& 2kk_1=-\frac{1}{2}(t+Q^2)
+\frac{S_p}{2\lambda_Y}\bigl(S_x(Q^2-t)+2tQ^2\bigr)+
\nonumber\\&& \qquad \qquad
+\frac{\sqrt{\lambda_{uw}}}{\lambda_{Y}}\cos\phi ,
\nonumber \\
u_0&=& 2kk_2=w_0+Q^2+t
\end{eqnarray}
with $S_p=S+X$ and 
\begin{equation}\label{lambdauw}
\lambda_{uw}=4W^2(Q^2(SX-M^2Q^2)-m^2\lambda_Y)(t-t_1)(t_2-t).
\end{equation}
Note that in massless approximation (for $m\rightarrow 0$) the BH cross section exactly coincides with results of \cite{BKM2002}.  
The following equations relating our notation to the notation of ref. \cite{BKM2002} (eqs. (30,32)) 
are valid: $u_0={\cal P}_2Q^2$, $w_0=-{\cal P}_1Q^2$, and (for $m\rightarrow 0$) $\lambda_{uw}=4Q^4S^2S_x^2K^2$ 

The three contributions to the total cross section (\ref{dGamma}) include BH process (BH amplitude squared), pure DVCS process (DVCS amplitude squared), and the contribution coming from the interference between BH and DVCS 
amplitudes. The first and the last terms are of specific interest because they provide main contributions for different 
observables: the interference term is the main contribution to observables where BH contribution is zero (i.e., the single spin 
asymmetry).

The interference term is proportional to:  
\begin{equation}\label{Mint2}
{\cal M}_{1\gamma}^2
=\bigl( {\cal M}_1^l+{\cal M}_2^l\bigr){\cal M}^{h\dagger}+{\cal M}^h\bigl( {\cal M}_1^l+{\cal M}_2^l\bigr)^\dagger .
\end{equation}
The BH matrix element is 
${\cal M}_{BH}={\cal M}_1^l+{\cal M}_2^l=e^3t^{-1}J^h_\mu J_{\mu \alpha}^{BH}(k_1,k_2,k)\epsilon^{\alpha }$ with 
\begin{eqnarray}
J_{\mu \alpha}^{BH}(k_1,k_2,k)=
J_{\mu \alpha}^{BH1}(k_1,k)+
J_{\mu \alpha}^{BH2}(k_2,k),
\nonumber\\[2mm]
J^h_\mu={\bar u}(p')\biggl(\gamma_{\mu}F_1+i\sigma_{\mu\nu}\frac{p_\nu '-p_\nu}{2M}F_2\biggr)u(p)
\end{eqnarray}
and
\begin{eqnarray}
&&
J_{\mu\alpha }^{BH}(k_1,k_2,k) 
\nonumber \\&&\;\;\;=
{\bar u}_2\Biggl [
 \gamma_\mu \frac{{\hat k}_1-{\hat k}+m}{-2kk_1}\gamma_\alpha
+ \gamma_\alpha \frac{{\hat k}_2+{\hat k}+m}{2kk_2}\gamma_\mu 
\Biggr ]u_1
\nonumber \\
&&=- 
{\bar u}_2\Biggl [\left(\frac {k_{1\alpha}}{kk_1}-\frac {k_{2\alpha}}{kk_2}\right)\gamma_\mu
-\frac{\gamma_\mu \hat{k}\gamma_\alpha}{2kk_1}
-\frac{\gamma_\alpha\hat{k}\gamma_\mu }{2kk_2}
\Biggr ]u_1.
\end{eqnarray}
Here ${\bar u}_2\equiv {\bar u}(k_2)$, ${u}_1\equiv {u}(k_1)$, and $\epsilon$ is the photon polarization vector. 
The matrix element ${\cal M}_{BH}$ corresponds to the graphs in Figure 
\ref{DVCSgraphs}a and \ref{DVCSgraphs}b.

The DVCS amplitude is calculated in \cite{BKM2002}. The set of explicit 
formulae (e.g. eqs. (1-8)) allows to present the DVCS amplitude in terms of covariant hadronic structures and the Compton form factors, that are calculable in QCD in leading and next-to-leading twists.  This representation is appropriate for RC calculation.  As a result the interference (\ref{Mint2}) has a form:
\begin{eqnarray}
&&{\cal M}_{1\gamma}^2(k_1,k_2,k)
\nonumber\\&&
=
\frac{64\pi^3\alpha^3}{tQ^2} 
\bigl(
J^h_\mu J^{BH}_{\mu\alpha }(k_1,k_2,k)
\bigl(T_{\alpha\nu}(k)J_\nu^{(0)}(k_1,k_2)\bigr)^{\dagger}
\nonumber\\&&
\qquad \;\;\;+
T_{\alpha\nu}(k)J_\nu^{(0)}(k_1,k_2)
\bigl(J^h_\mu J^{BH}_{\mu\alpha }(k_1,k_2,k)\bigr)^{\dagger}
\bigr).
\label{Mint2b}
\end{eqnarray}
Here $J_\mu^{(0)}(k_1,k_2)={\bar u}_2 \gamma_\mu{u}_1$ and $T_{\nu\mu}(k)$ are defined by eqs. (1-8) of \cite{BKM2002}.

We calculate the cross section in the BMK approximation. In this approximation the cross section is represented through the sum over the finite number of terms  reflecting the $\phi$-dependence. Respective coefficients are referred as the Fourier coefficients. They are calculated in the leading approximation at $M^2$ and $t$ simultaneously going to zero. We present them in the form appropriate for further RC calculation.

The most important observable quantity is the beam spin asymmetry:
\begin{equation}\label{eq11}
A_{1\gamma}={\sigma_I^p\over \sigma_{BH}^u +\sigma_I^u},
\end{equation}
where $\sigma^u_{BH}$ is the BH cross section of unpolarized electrons and protons, and  $\sigma_I^{u,p}$ are unpolarized and spin-dependent parts of the cross section resulted from the interference of the BH and DVCS amplitudes.  
In BMK approximation the BH cross section is expressed as
\begin{equation}
\sigma_{BH}^u=\frac{f}{{\cal P}_1{\cal P}_2}(c_0^{BH}+c_1^{BH}\cos\phi
+c_2^{BH}\cos 2\phi),
\end{equation}
where $f=\alpha^3 S/(8\pi y^2t\lambda_Y^{1/2})$. 
There are four and two non-zero coefficients for unpolarized and polarized beam, respectively:
\begin{eqnarray}\label{scc0}
\sigma_I^{u}&=&\frac{f}{{\cal P}_1{\cal P}_2}(c_0^I+c_1^I\cos\phi+c_2^I\cos 2\phi+c_3^I\cos 3\phi),
\nonumber \\
\sigma_I^{p}&=&\frac{f}{{\cal P}_1{\cal P}_2}(s_1^I\sin\phi+s_2^I\sin 2\phi).
\end{eqnarray}
The corresponding Fourier coefficients $c_i$ and $s_i$ are calculated in \cite{BKM2002}.

\section{\label{RC}RC cross section}

\begin{figure}[t]\centering
\scalebox{0.24}{\includegraphics{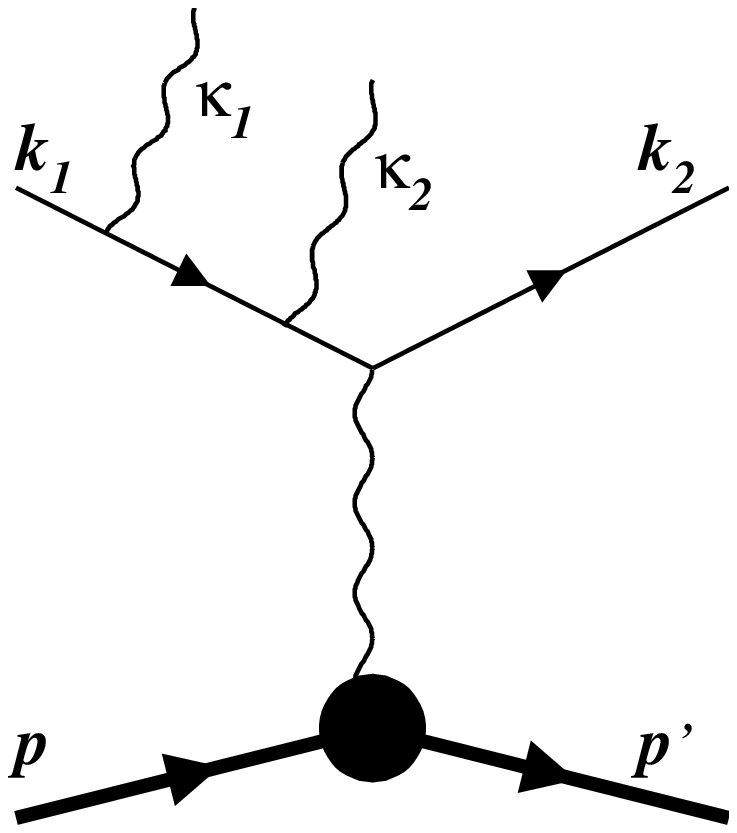}}
\hspace{0.25cm}
\scalebox{0.24}{\includegraphics{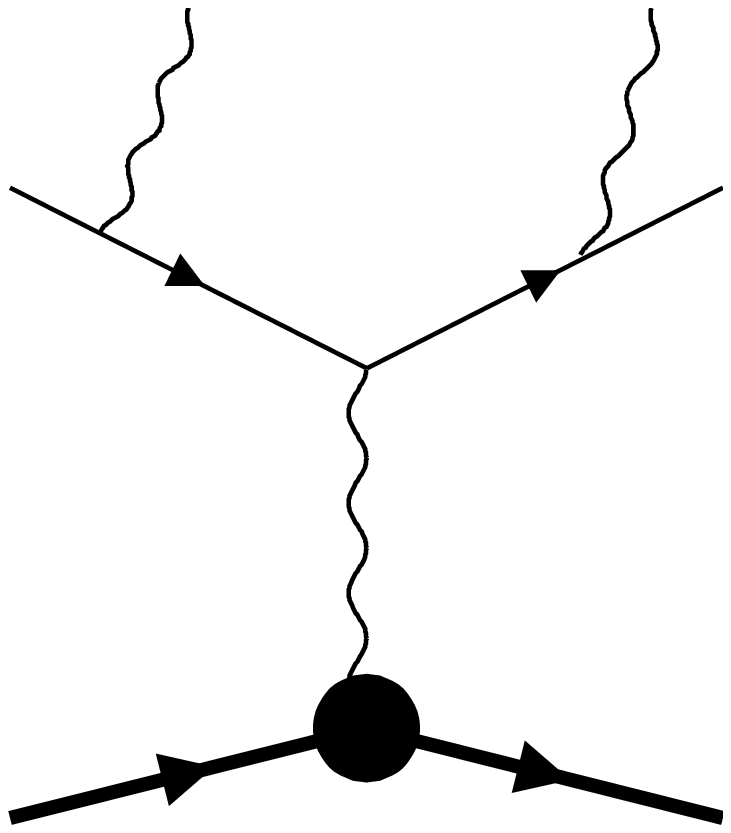}}
\hspace{0.25cm}
\scalebox{0.24}{\includegraphics{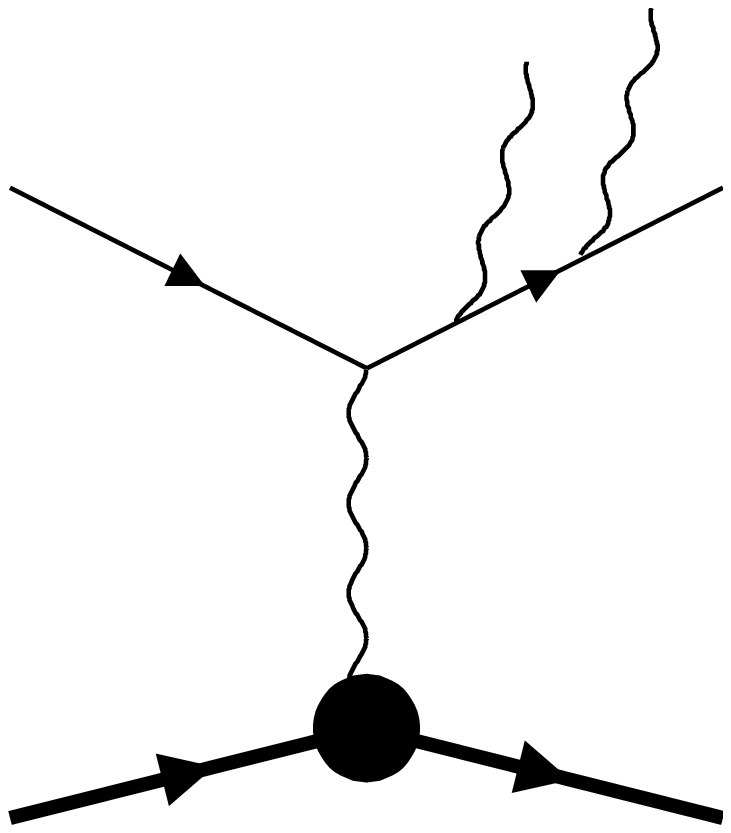}}
\\
{\bf a) \hspace{1.8cm} b) \hspace{1.8cm} c)}
\\[2mm]
\scalebox{0.24}{\includegraphics{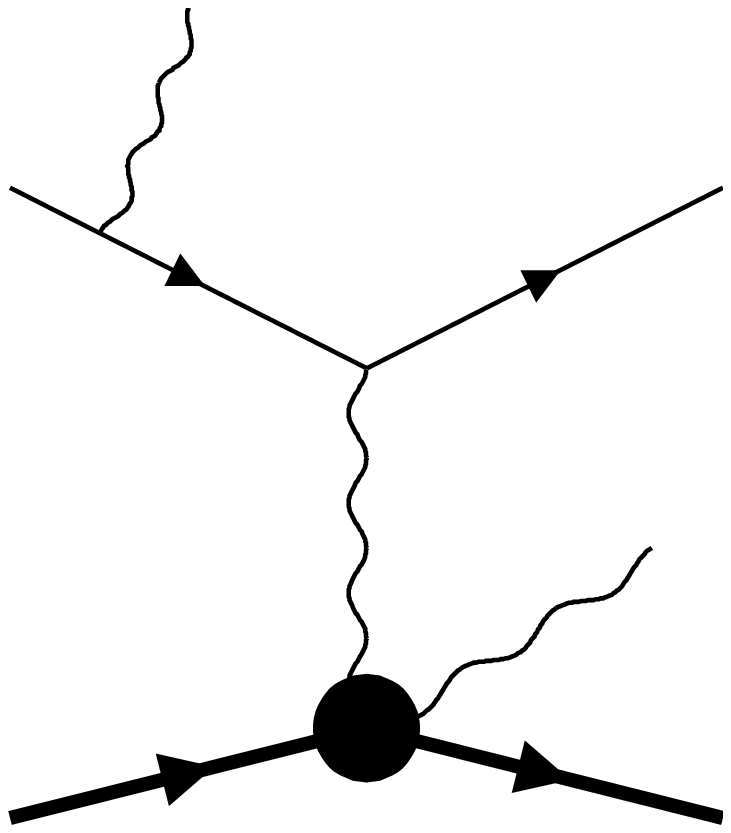}}
\hspace{0.25cm}
\scalebox{0.24}{\includegraphics{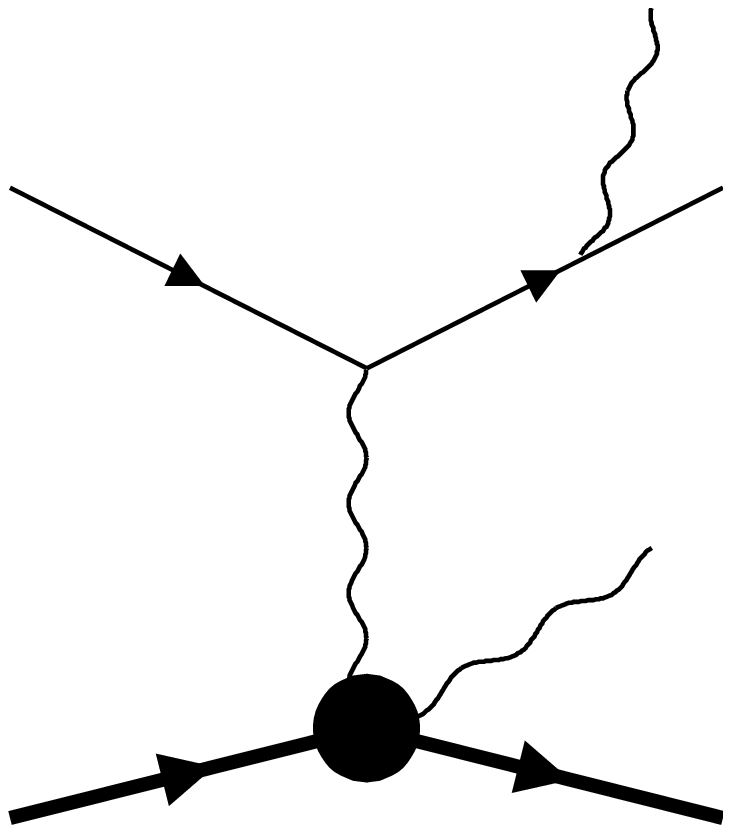}}
\\
{\bf d) \hspace{1.8cm} e)}
\caption{\label{Twoggraphs}Feynman graphs of two real photon emission.}
 \end{figure}

The cross section of two photon emission, i.e., the process
\begin{equation}\label{twogammaprocess}
e(k_1)+p(p)\longrightarrow e'(k_2)+p'(p')+\gamma(\kappa_1)+\gamma(\kappa_2),
\end{equation}
is
\begin{eqnarray}
d\sigma&=&\frac{1}{4S} {\cal M}_{2\gamma}^2 d\Gamma,
\label{xs2g}
\end{eqnarray}
where 
\begin{eqnarray}
d\Gamma&=&\frac 1{(2\pi )^8}
\frac{d^3k_2}{2E_2}
\frac{d^3p'}{2p'_0}
\frac{d^3\kappa_1}{2\omega_1}
\frac{d^3\kappa_2}{2\omega_2}
\nonumber \\&&\times
\delta^4(k_1+p-k_2-p'-\kappa_1-\kappa_2)
\end{eqnarray}
and the additional factor of 2 in the denominator is because there are two identical particles (photons) in the final state. 
The matrix element squared of the process with two real photons in the final state has the contribution of pure leptonic correction 
(shown in Figure \ref{Twoggraphs}~a-c and discussed in \cite{AkushevichIlyichev2012}) and the contribution of the interference between lepton and hadron 
emissions:
\begin{eqnarray}\label{m2sq}
{\cal M}_{2\gamma}^2 &=&\sum_{i=1}^6\bigl[ 
{\cal M}_i^{ll}\bigl( {\cal M}_1^{lh}+{\cal M}_2^{lh}\bigr)^{\dagger}
\nonumber \\&&
+\bigl( {\cal M}_1^{lh}+{\cal M}_2^{lh}\bigr) {\cal M}_i^{ll\;\dagger}\bigr].
\end{eqnarray}

Six matrix elements of the process with emission of the additional photon correspond 
to graphs in Figure \ref{Twoggraphs}a-c: 
${\cal M}^{ll}_{1-6}=e^4t^{-1}J_\mu^h J_{1-6,\mu}$. 
The quantities $J_{1-6,\mu}$ are defined by eq. (22) of \cite{AkushevichIlyichev2012}:
\begin{eqnarray}
J_{1\mu}&=& 
{\bar u}_2 
\gamma_\mu 
\frac{{\hat k}_1-{\hat \kappa}+m}{-2\kappa k_1+V^2}
{\hat \epsilon}_2
\frac{{\hat k}_1-{\hat \kappa}_1+m}{-2k_1\kappa_1}
{\hat \epsilon}_1
u_1,
\nonumber \\
J_{2\mu}&=& 
{\bar u}_2 
\gamma_\mu 
\frac{{\hat k}_1-{\hat \kappa}+m}{-2\kappa k_1+V^2}
{\hat \epsilon}_1
\frac{{\hat k}_1-{\hat \kappa}_2+m}
{-2k_1\kappa_2}{\hat \epsilon}_2
u_1,
\nonumber \\
J_{3\mu}&=& 
{\bar u}_2 
{\hat \epsilon}_2
\frac{{\hat k}_2+{\hat \kappa_2}+m}{2k_2\kappa_2}
{\hat \epsilon}_1
\frac{{\hat k}_2+{\hat \kappa }+m}{2\kappa k_2+V^2}
\gamma_\mu 
u_1,
\nonumber \\
J_{4\mu}&=& 
{\bar u}_2 
{\hat \epsilon}_1
\frac{{\hat k}_2+{\hat \kappa_1}+m}{2k_2\kappa_1}
{\hat \epsilon}_2
\frac{{\hat k}_2+{\hat \kappa }+m}{2\kappa k_2+V^2}
\gamma_\mu 
u_1,
\nonumber \\
J_{5\mu}&=& 
{\bar u}_2 
{\hat \epsilon}_1
\frac{{\hat k}_2+{\hat \kappa_1}+m}{2k_2\kappa_1}
\gamma_\mu 
\frac{{\hat k}_1-{\hat \kappa}_2+m}{-2k_1\kappa_2}
{\hat \epsilon}_2
u_1,
\nonumber \\
J_{6\mu}&=& 
{\bar u}_2 
{\hat \epsilon}_2
\frac{{\hat k}_2+{\hat \kappa_2}+m}{2k_2\kappa_2}
\gamma_\mu 
\frac{{\hat k}_1-{\hat \kappa}_1+m}{-2k_1\kappa_1}
{\hat \epsilon}_1
u_1,
\end{eqnarray}
where $V^2=\kappa^2=(\kappa_1+\kappa_2)^2$ is the missing mass squared.

Matrix elements with emissions of one photon from lepton line and one photon from hadron line (Figure \ref{Twoggraphs}d, e) are
\begin{eqnarray}
 {\cal  M}^{lh}_1+{\cal  M}^{lh}_2&=&e^4\epsilon_1^\alpha\epsilon_2^\beta
\Biggl(\frac{J_{\mu\alpha}^{BH}(k_1,k_2,\kappa _1)T_{\beta\mu}(\kappa _2) }{Q^2+2q\kappa_1}
\nonumber\\&&\qquad \;
+\frac{J_{\mu\beta}^{BH}(k_1,k_2,\kappa _2)T_{\alpha\mu}(\kappa _1)}{Q^2+2q\kappa_2}
\Biggr)
. 
\end{eqnarray}

The matrix element squares
(\ref{m2sq}) has four terms with denominators containing $\kappa_{1,2}k_1$ ($s$-peak) and $\kappa_{1,2}k_2$ 
($p$-peak):
\begin{eqnarray}\label{fourterms}
{\cal M}_{2\gamma}^2 ={\cal M}_{1s}^2+{\cal M}_{1p}^2+{\cal M}_{2s}^2+{\cal M}_{2p}^2,
\end{eqnarray}
where indices correspond to the unobserved photon, e.g., $1s$ means that the photon with momentum $\kappa_1$ is unobserved and 
in
the $s$-peak. 
Just these four terms contribute to the cross section in the leading approximation. 
Each of them (i.e., $1/k_1\kappa_1$, $1/k_1\kappa_2$, $1/k_2\kappa_1$, or $1/k_2\kappa_2$) contains 
the first order pole which can be extracted if to put vectors $\kappa_1$ and $\kappa_2$ in the peak and use $m\rightarrow 0$ in the coefficient at respective pole. Practically the terms are calculated by using the following substitution:
$\kappa_1=(1-z_1)k_1$, $\kappa_1=(1/z_2-1)k_2$, 
$\kappa_2=(1-z_1)k_1$ and
$\kappa_2=(1/z_2-1)k_2$ for ${\cal M}_{1s}^2$, ${\cal M}_{1p}^2$, ${\cal M}_{2s}^2$ and ${\cal M}_{2p}^2$ respectively. The use of these formulae means putting the angular components of the vectors $\kappa_1$ and $\kappa_2$ to be equal of respective angular components of vectors $k_1$ and 
$k_2$ in numerators of all terms in right hand side of (\ref{fourterms})
${\cal M}^2_{1,2s,p}$, keeping the last component (i.e., energy of $\kappa_1$ and $\kappa_2$) unfixed. The variables $z_{1,2}$ represent the energy-related components of the vectors and can be related to $V^2$ as 
\begin{equation}\label{z1z2vv2}
 z_1=1-{V^2\over w}, \qquad z_2={u\over u+V^2},
 \end{equation}
where $w=2k_1(p+q-p^\prime)$ and $u=2k_2(p+q-p^\prime)$.

The calculation of ${\cal M}_{1s}^2$ is similar to that considered 
in \cite{{AkushevichIlyichev2012}} but there are new technical issues because of different structure of matrix element squared. Only 
${\cal M}_1^{ll}$, ${\cal M}_6^{ll}$, and ${\cal M}_1^{lh}$ can have the pole $1/k_1\kappa_1$ through contributions from $J_{1\mu} $, $J_{1\mu}$, and $J^{BH1}_{\mu\alpha}$ that are reduced to 
\begin{eqnarray}
J_{1\mu}&\approx &\frac{k_1\epsilon_1}{2(k_1\kappa _2)(k_1\kappa_1)}
{\bar u}_2 \gamma_\mu (z_1 {\hat k}_1-{\hat \kappa}_2){\hat \epsilon}_2
u_1 
\nonumber \\
J_{6\mu}&\approx &-\frac{z_1k_1\epsilon_1}{2(k_2\kappa _2)(k_1\kappa_1)}
{\bar u}_2 {\hat \epsilon}_2({\hat k}_2+{\hat \kappa}_2)\gamma_\mu 
u_1 ,
\nonumber \\
J_{\mu\alpha }^{BH1}(k_1,\kappa _1)&\approx &-\frac{z_1\;\;k_{1\;\alpha}}{k_1\kappa_1}
{\bar u}_2 \gamma_\mu u_1 ,
\label{Japp}
\end{eqnarray}

The convolution of the sum ${\cal M}^{ll}_1+{\cal M}^{ll}_6$ with ${\cal M}^{lh}_2$ 
contains the infrared divergence,
\begin{eqnarray}
({\cal M}^{ll}_1+{\cal M}^{ll}_6){\cal M}^{lh\dagger}_2+
{\cal M}^{lh}_2
({\cal M}^{ll}_1+{\cal M}^{ll}_6)^\dagger
\nonumber\\
={4\pi \alpha \over (1- z_1)\kappa _1 k_1} 
{\cal M}^2_{1\gamma }(z_1 k_1,k_2,\kappa_2),
\end{eqnarray}
at $z_1\to 1$, while the convolution of this sum with ${\cal M}^{lh}_1$,
\begin{eqnarray}
({\cal M}^{ll}_1+{\cal M}^{ll}_6){\cal M}^{lh\dagger}_1+
{\cal M}^{lh}_1
({\cal M}^{ll}_1+{\cal M}^{ll}_6)^\dagger
\nonumber\\
={4\pi \alpha (1-z_1)\over z_1\kappa _1 k_1} 
{\cal M}^2_{1\gamma }(z_1 k_1,k_2,\kappa_2),
\end{eqnarray}
does not.

The convolution of the other terms ${\cal M}^{ll}_{2-5}$ with ${\cal M}^{lh}_1$ results in the term also containing the infrared divergence:
\begin{eqnarray}
\sum\limits_{i=2}^5
[{\cal M}^{ll}_i{\cal M}^{lh\dagger}_1
+
{\cal M}^{lh}_1
{\cal M}^{ll\dagger}_i ]
=
{4\pi \alpha {\cal M}_{1\gamma}^2(z_1 k_1,k_2,\kappa_2)\over (1- z_1)\kappa _1 k_1}. 
\nonumber\\
\end{eqnarray}
All other convolutions do not have contributions to the cross sections in the leading log approximation.

%

The resulting expressions for the terms in (\ref{fourterms}) are:
\begin{eqnarray}
{\cal M}_{1s}^2 
&=&{4\pi \alpha \over \kappa_1 k_1}{(1+z_1^2)\over z_1(1- z_1)}
{\cal M}^2_{1\gamma }(z_1 k_1,k_2,\kappa_2),
\nonumber \\
{\cal M}_{1p}^2 
&=&{4\pi \alpha \over \kappa_1 k_2}{(1+z_2^2)\over (1- z_2)}
{\cal M}^2_{1\gamma }\biggl(k_1,{k_2 \over z_2},\kappa_2 \biggr),
\nonumber \\
{\cal M}_{2s}^2 
&=&{4\pi \alpha \over \kappa_2 k_1}{(1+z_1^2)\over z_1(1- z_1)}
{\cal M}^2_{1\gamma }(z_1 k_1,k_2,\kappa_1),
\nonumber \\
{\cal M}_{2p}^2 
&=&{4\pi \alpha \over \kappa_2 k_2}{(1+z_2^2)\over (1- z_2)}
{\cal M}^2_{1\gamma }\biggl(k_1,{k_2 \over z_2},\kappa_1\biggr).
\end{eqnarray}
The integration over angular variables results in:
\begin{eqnarray}
\int \frac {d\Gamma}{\kappa _1k_1}=
d\Gamma_0
\frac{L}{8\pi^2w}dV^2,
\nonumber \\
\int \frac {d\Gamma}{\kappa _1k_2}=
d\Gamma_0
\frac{L}{8\pi^2u}dV^2.
\end{eqnarray}

Thus, the matrix elements squared of the process with two real photons in the final state is expressed in terms of BH/DVCS matrix element squared in  the same way as for the BH process \cite{AkushevichIlyichev2012}. The phase space parameterization is also independent on the dynamics of the process and its parameterization obtained for the pure BH process is applicable for the contribution from the BH and DVCS interference as well \cite{AkushevichIlyichev2012}. Therefore, the cross section for radiative correction to the interference of BH and DVCS amplitudes keeps the same 
form:
\begin{eqnarray}\label{iniint}
&&\sigma _{s}(S,x,Q^2,t,\phi)=\frac{\alpha}{2\pi }L
\int\limits_{z_{1}^m}^1 dz_1\frac {1+z_1^2}{1-z_1} 
K_s(z_1)\sigma _{1\gamma}(z_1),
\nonumber\\&&
\sigma _{p}(s,x,Q^2,t,\phi)=\frac{\alpha}{2\pi }L
\int\limits_{z_{2}^m}^1 dz_2\frac {1+z_2^2}{1-z_2}
K_p(z_2)\sigma _{1\gamma}(z_2).
\nonumber\\&&
\end{eqnarray}
Here $z_{1,2}$ in brackets means that the cross section needs to be taken in a shifted kinematics, i.e., 
\begin{eqnarray}
\sigma _{1\gamma}(z_1)&=&\sigma _{1\gamma}(z_1S,x_s,z_1Q^2,t,{\bar\phi}_s), \nonumber
\\
\sigma _{1\gamma}(z_2)&=&\sigma _{1\gamma}(S,x_p,z_2^{-1}Q^2,t,{\bar\phi}_p)
\label{sz1z2}
\end{eqnarray}
and 
\begin{equation}
K_s(z_1)={x_s^2\sin\theta_s '   \over x^2{{\cal D}_{0s}^{1/2}} },
\quad
 K_p(z_2)={x_p^2\sin\theta_p '  \over z_2x^2{{\cal D}_{0p}^{1/2}}},
\end{equation}
where $x_s=z_1Q^2/(z_1S-X)$ and $x_p=Q^2/(z_2S-X)$ are Bjorken $x$ in shifted kinematics;
${\cal D}_{0}$, $\sin\theta '$ and $\bar\phi$ are given in \cite{AkushevichIlyichev2012} by eqs.~(38) and (40);
the subscript ($s$ or $p$) explicitly indicates the type of kinematics for that these quantities have to be calculated.

\begin{figure}[t]
\centering
\scalebox{0.2}{\includegraphics{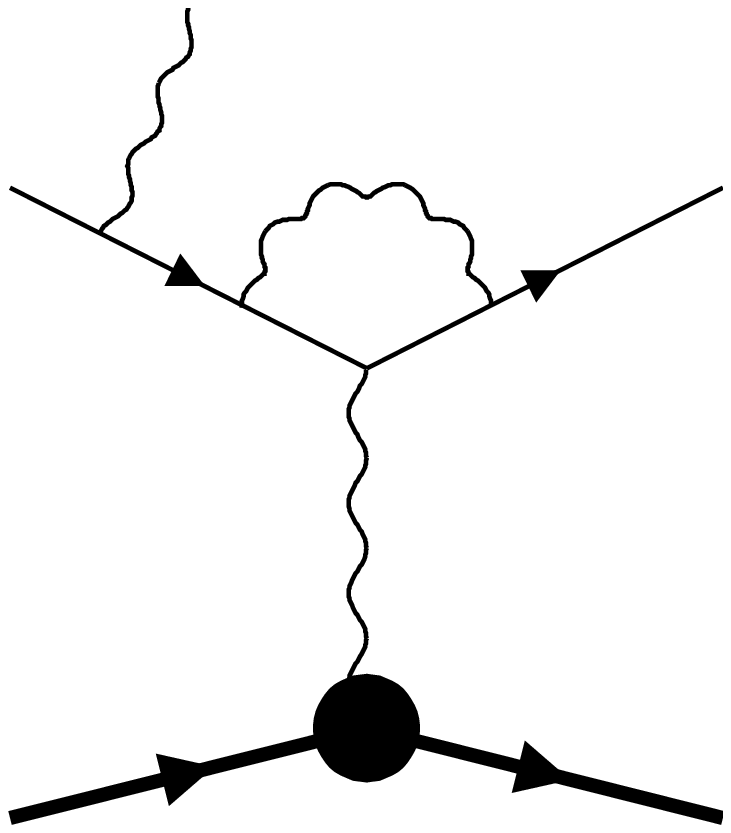}}
\hspace{0.25cm}
\scalebox{0.2}{\includegraphics{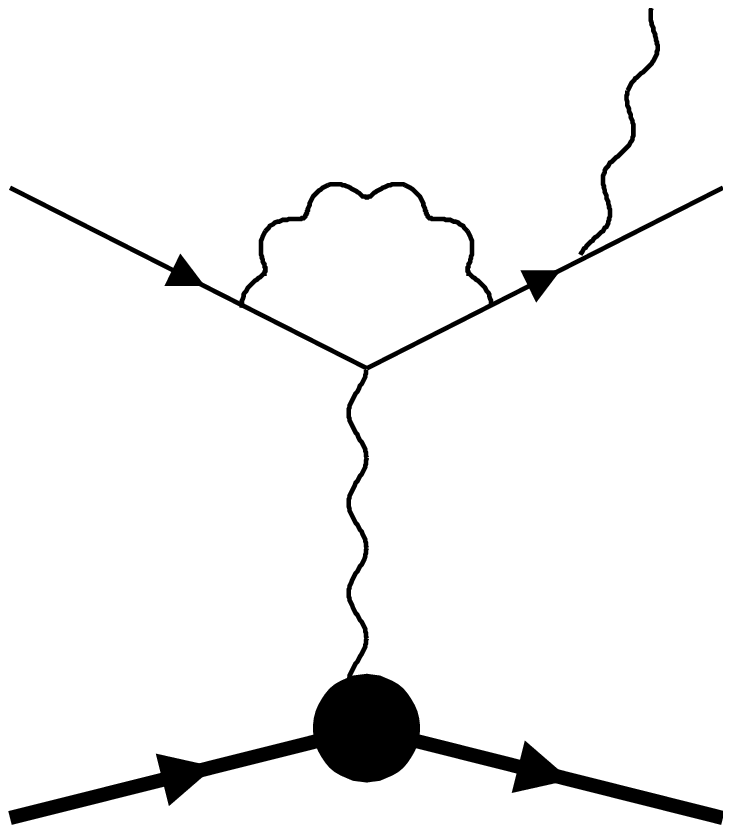}}
\hspace{0.25cm}
\scalebox{0.2}{\includegraphics{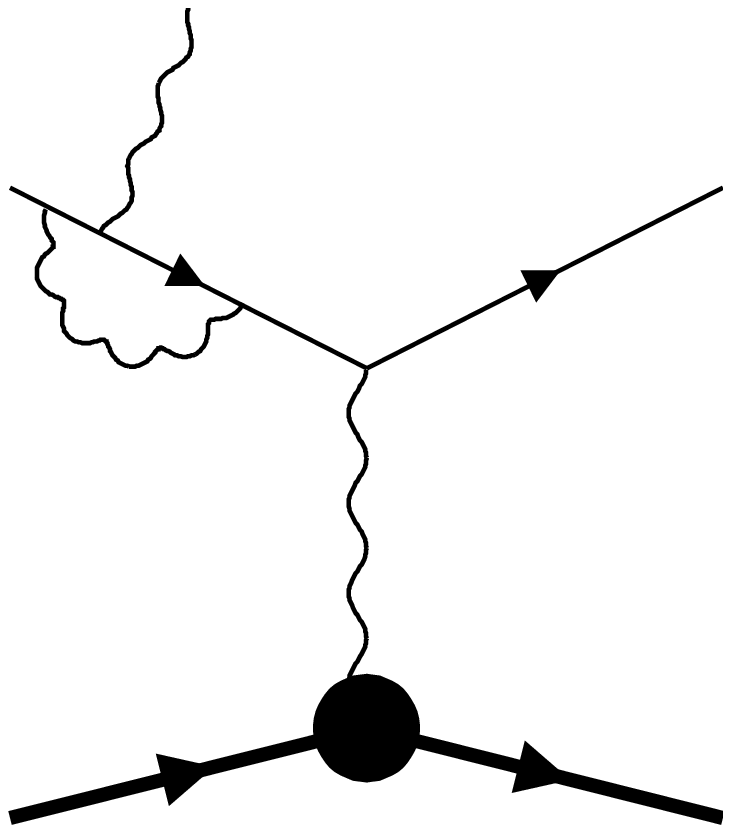}}
\hspace{0.25cm}
\scalebox{0.2}{\includegraphics{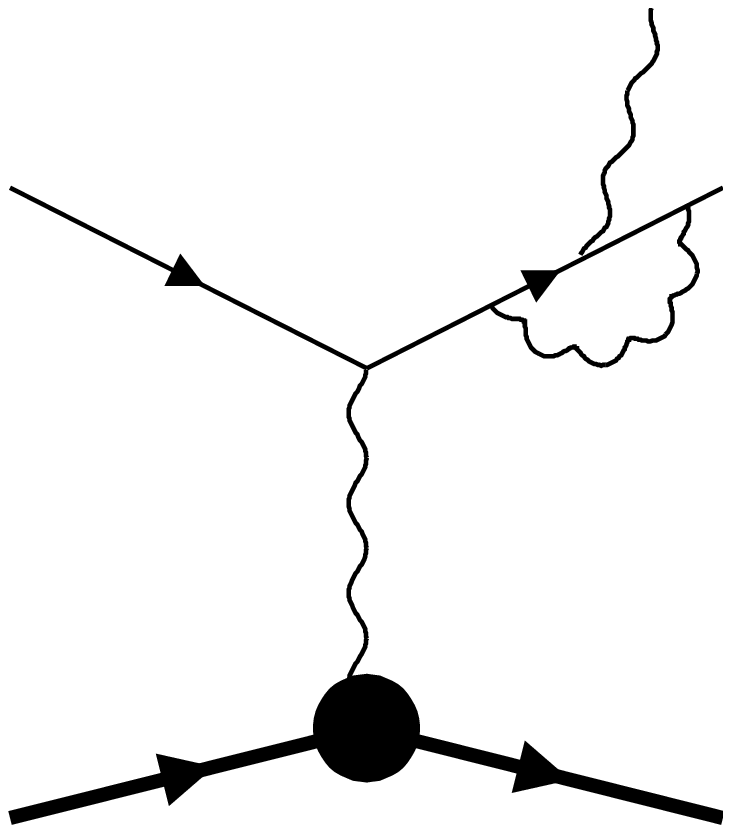}}
\\[-0.1cm]
{\bf \small a) \hspace{1.52cm} b) \hspace{1.52cm} c)\hspace{1.52cm} d)}
\\[0.1cm]
\scalebox{0.2}{\includegraphics{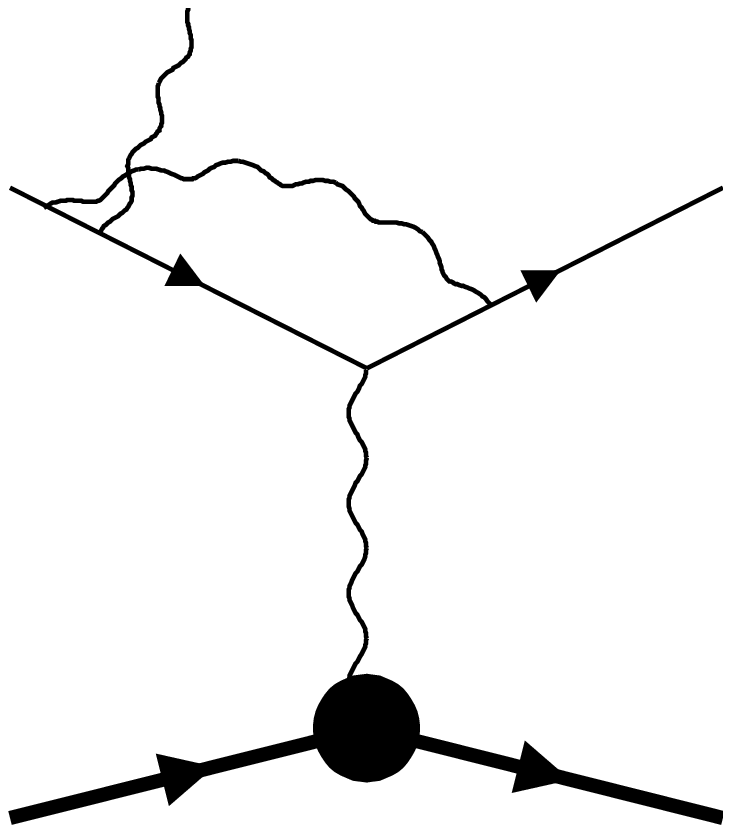}}
\hspace{0.25cm}
\scalebox{0.2}{\includegraphics{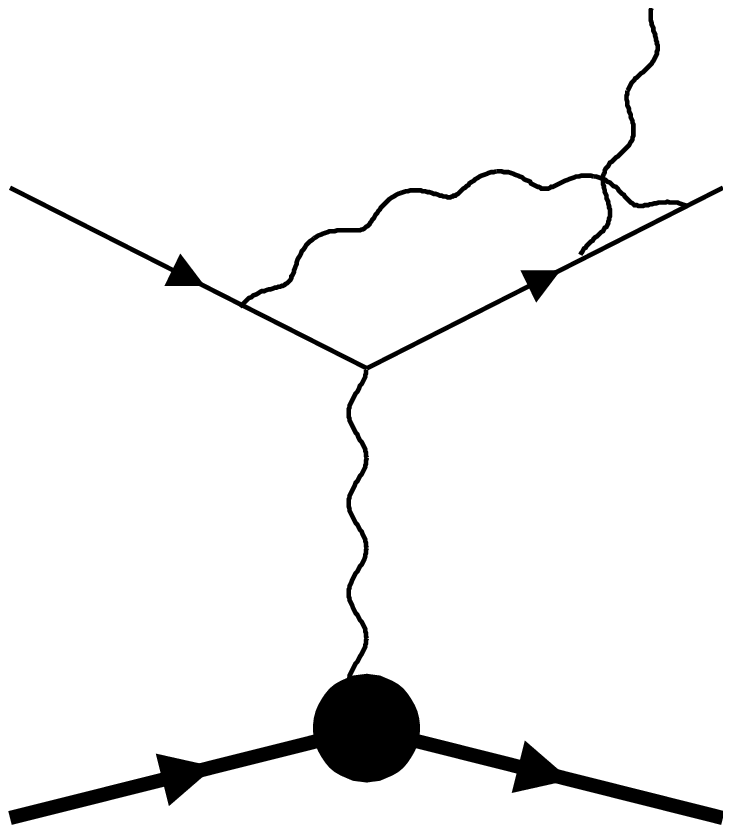}}
\hspace{0.25cm}
\scalebox{0.2}{\includegraphics{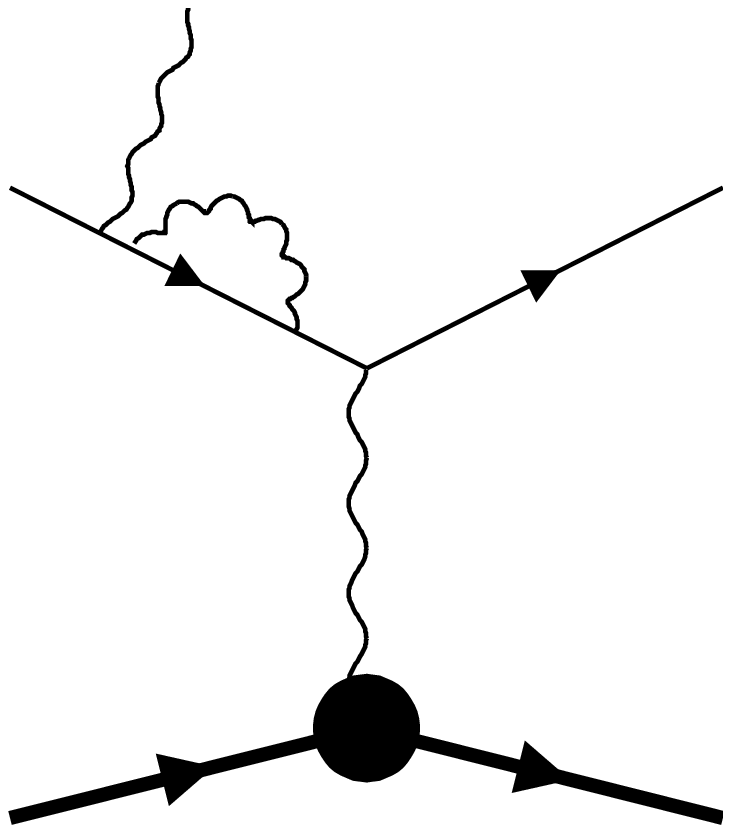}}
\hspace{0.25cm}
\scalebox{0.2}{\includegraphics{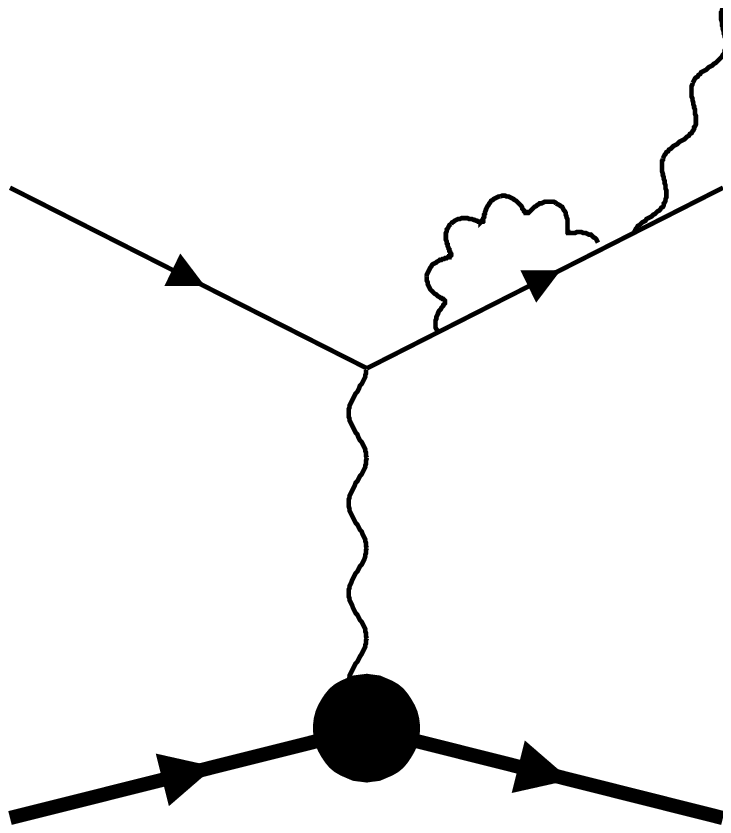}}
\\[-0.1cm]
{\bf \small e) \hspace{1.52cm} f) \hspace{1.52cm} g)\hspace{1.52cm} h)}
\\[0.1cm]
\scalebox{0.2}{\includegraphics{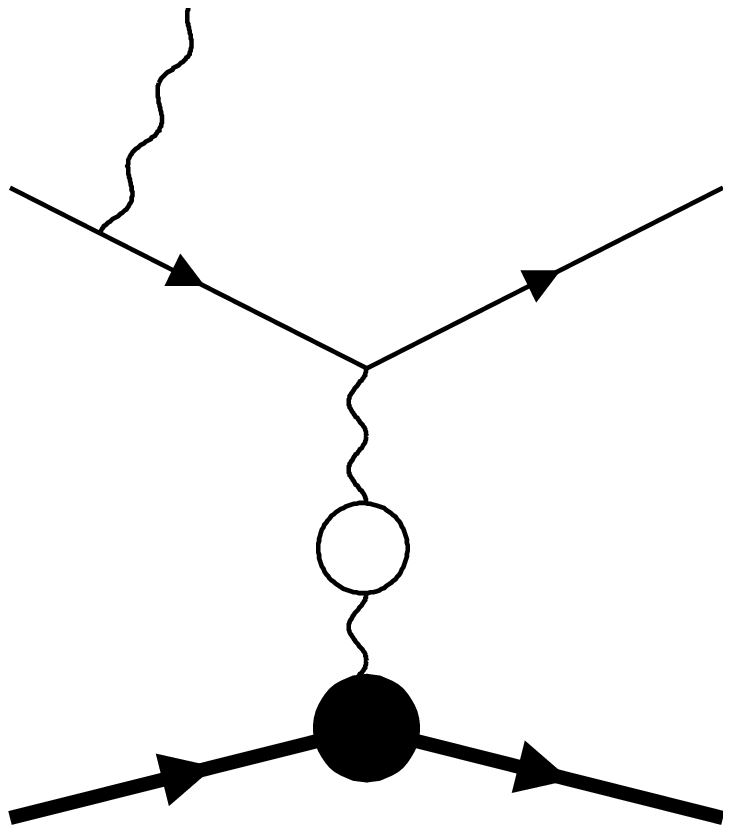}}
\hspace{0.25cm}
\scalebox{0.2}{\includegraphics{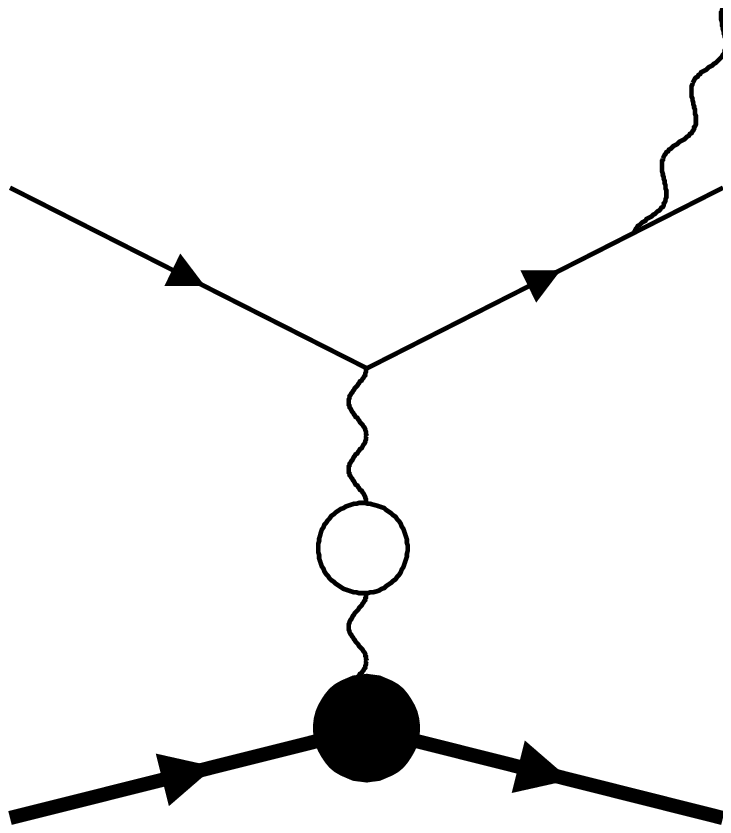}}
\hspace{0.25cm}
\scalebox{0.2}{\includegraphics{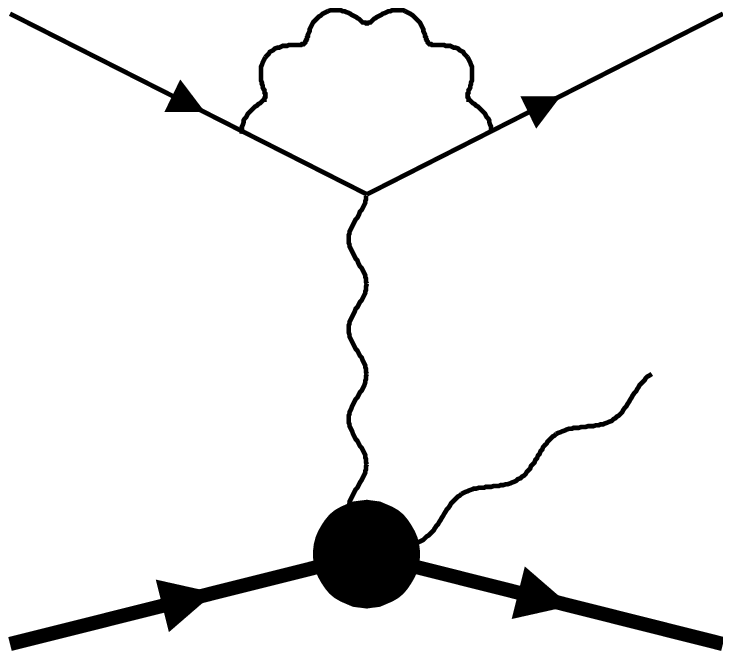}}
\hspace{0.25cm}
\scalebox{0.2}{\includegraphics{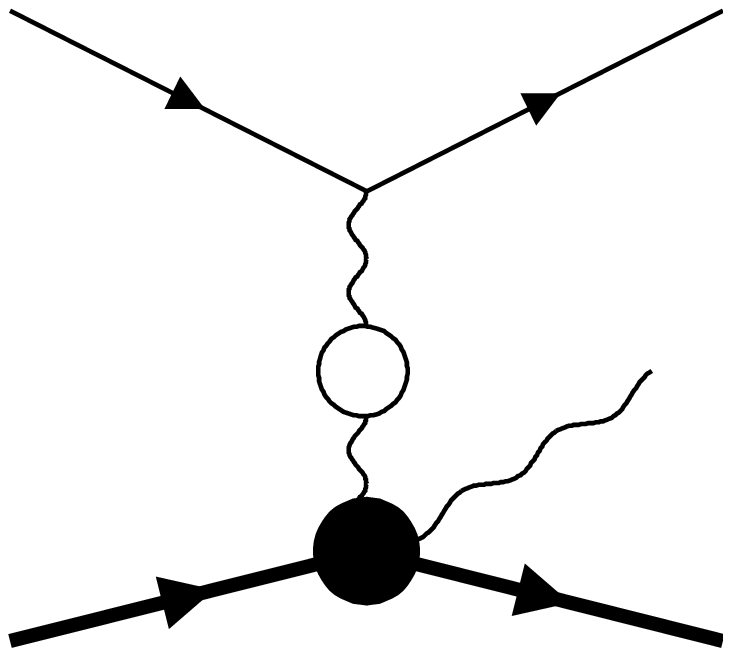}}
\\[-0.1cm]
{\bf \small i) \hspace{1.52cm} j) \hspace{1.52cm} k)\hspace{1.52cm} l)}
\caption{\label{VVgraphs}Feynman graphs of one-loop effects for the BH cross section.}
 \end{figure}

 The lower limits of integration in (\ref{iniint})
 are defined by the substitutional in 
eq. (\ref{z1z2vv2}) the maximal missing mass squared 
\begin{equation}
V_{max}^2=(2M)^{-1}(\sqrt{\lambda_Y}\sqrt{\lambda_t}+S_xt)-Q^2+t
\end{equation} 
with $\lambda_t=t(t-4M^2)$ and reads as:
\begin{eqnarray}
z_1^m=\frac{t(X-2M^2)+\sqrt{\lambda_t/\lambda_Y}(XS_x-2M^2Q^2)}{St-2M^2Q^2+\sqrt{\lambda_t/\lambda_Y}(SS_x+2M^2Q^2)},
\nonumber\\
z_2^m=\frac{Xt-2M^2Q^2+\sqrt{\lambda_t/\lambda_Y}(XS_x-2M^2Q^2)}{t(S+2M^2)+\sqrt{\lambda_t/\lambda_Y}(SS_x+2M^2Q^2)}.
\label{z12m}
\end{eqnarray} 

\begin{figure}\centering
\scalebox{0.48}{\includegraphics{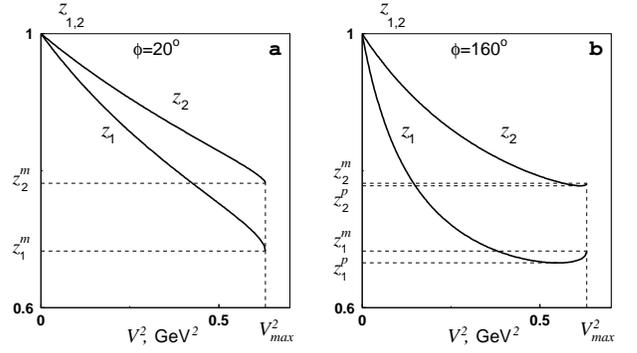}}
\caption{\label{zvv2rel}The dependence of $z_1$ and $z_2$ on $V^2$ for $\protect\cos(\protect\phi)>0$ (a) and 
$\protect\cos(\protect\phi)<0$ (b). Other kinematical variables used for this example were 
$x$=0.175,$Q^2$=1GeV$^2$, $t$=-0.1GeV$^2$, and $E_{beam}$=5.75GeV.
In plot b), the curves $z_{1,2}(V^2)$ cross the lines $z_{1,2}=z_{1,2}^m$ at $V^2=V^2_{1\;s,p}$ 
and reach their minimum values at $V^2=V^2_{2\;s,p}$ such that $V^2_{1\;s,p}<V^2_{2\;s,p}<V^2_{max}$. The explicit expressions for $z_{1,2}^p$
and $V^2_{1,2\;s,p}$ are given in Appendix. }
 \end{figure}

The relationship between $z_{1,2}$ and $V^2$ is illustrated in Figure \ref{zvv2rel}. In the most cases the relation between them is unambiguous as shown in Figure \ref{zvv2rel}a. 
However there are situations (for $\cos\phi<0$) when the curves $z_{1,2}(V^2)$ have a minimum (Figure \ref{zvv2rel}b). In this case an additional contribution reflecting the area between two points of intersections between the curve $z_{1,2}(V^2)$ and the line $z_{1,2}=z^m_{1,2}$ can to be presented as a separate contribution to the cross section. The explicit form of this contribution is given in Appendix. 

 The integrals in (\ref{iniint}) are divergent at upper integration limit because of the infrared divergence, that is canceled by adding the contribution of  loops and soft photon emission \cite{ByKuTo2008PRC} represented by the Feynman graphs in Fig. \ref{VVgraphs}a-h,k. The result for infrared free contribution is
 \begin{equation}
{\alpha \over \pi} \bigl( \delta_{inf}+\delta_{fin}\bigr) \sigma _{1\gamma}+ \sigma_s^F+\sigma_p^F+\sigma_{add},
\end{equation}
where 
\begin{eqnarray}
\delta_{fin}&=&\frac{L}{4}\bigl(z_1^m(2+z_1^m)+z_2^m(2+z_2^m)\bigr), 
\nonumber
\\
\delta_{inf}&=&L\bigl( \log(1-z_1^m)+\log(1-z_2^m)\bigr)
\label{twodeltas}
\end{eqnarray}
and 
\begin{eqnarray} \label{sigmaF}
\sigma _s^F&=&\frac{\alpha L}{2\pi }
\int\limits_{z_1^m}^1 dz_1 \frac {1+z_1^2}{1-z_1}
\biggl(
K_s(z_1)
\;\sigma _{1\gamma}(z_1)
-
\sigma _{1\gamma}\biggr),
\nonumber
\\
\sigma _p^F&=&\frac{\alpha L}{2\pi }
\int\limits_{z_2^m}^1 dz_2 \frac {1+z_2^2}{1-z_2} 
\biggl( 
K_p(z_2)
\;\sigma _{1\gamma}(z_2)
-
\sigma _{1\gamma}\biggr)
.
\label{1l}
\end{eqnarray}
The experimental cuts on missing mass squared $V_{cut}^2$ or maximal photon energy can be incorporated 
by the following replacements:
\begin{eqnarray}
K_{s,p}(z_{1,2})\rightarrow \theta(z_{1,2}-z_{1,2}^c) K_{s,p}(z_{1,2}).
\end{eqnarray}
Here $z_{1,2}^c$ reflect the restrictions on the energy of hard photon or missing mass squared. The relation between these variables 
are given in eq.~(\ref{z12c}) of Appendix.

The total lowest order RC 
is
\begin{equation}\label{eqeq30}
\sigma _{RC}= \frac{\alpha}{\pi}\bigl( \delta_{vac}+\delta_{inf}+\delta_{fin}\bigr) \sigma _{1\gamma} + \sigma_{s}^F+\sigma_p^F+\sigma_{add}.
\end{equation}
Here $\delta_{vac}$ reflects the contribution of vacuum polarization, i.e., the Feynman graphs in Fig. \ref{VVgraphs}i,j,l. Specifically, $ \Pi (t)=\alpha/(2\pi)\delta_{vac}$  and $\delta_{vac}$ is the contribution of vacuum polarization by leptons and hadrons calculated as in \cite{AKSh1994} (see eq. (21) and discussion before eq. (20)). Formally the expression for the observed cross section coincides with the cross section for the BH process obtained in \cite{AkushevichIlyichev2012} (expression (48)). The higher order corrections can be included in the style of (51) or (52) of ref. \cite{AkushevichIlyichev2012}.

Behavior of the cross section for $t$ close to kinematical bounds (i.e., in the region where $t\sim t_1$ and $t\sim t_2$) deserves special attention. The quantity $\delta_{inf}$ in (\ref{twodeltas}) become infinite when $t \rightarrow t_1$ or $t \rightarrow t_2$. In this limit $z_1^m=1$ and $z_2^m=1$. The source of occurrence of the divergence is known \cite{YennieFrautschiSuura1961}. The divergence is canceled by taking into account multiple soft photon emission. We follow the so-called exponentiation procedure suggested in \cite{Shumeiko}:
\begin{eqnarray}\label{exponentiation}
&&\Bigl( 1 + \frac{\alpha}{\pi}\bigl( \delta_{vac}+\delta_{inf}+\delta_{fin}\bigr)\Bigr)
\rightarrow \\
&&\qquad \qquad \exp\bigl(\frac{\alpha}{\pi} \delta_{inf}\bigr)\Bigl( 1 + \frac{\alpha}{\pi}\bigl( \delta_{vac}+\delta_{fin}\bigr)\Bigr), \nonumber
\end{eqnarray}
such that the observed cross section becomes:
\begin{eqnarray}
\sigma _{obs}&=&\exp\bigl(\frac{\alpha}{\pi} \delta_{inf}\bigr)\Bigl( 1 + \frac{\alpha}{\pi}\bigl( \delta_{vac}+\delta_{fin}\bigr)\Bigr)
 \sigma _{1\gamma} 
\nonumber \\&&
 + \sigma_{s}^F+\sigma_p^F+\sigma_{add}.
\end{eqnarray}
After this procedure the observed cross section vanishes at the kinematical bounds on $t$.

This result allows us to construct a Monte Carlo generator of the events with one or two photons in the final state. To have an opportunity to simulate the specific contributions we must  represent the observed cross section as a sum of positively definite contributions. Because of the last terms in (\ref{sigmaF}), i.e., the terms containing $\sigma _{1\gamma}$, the contributions $\sigma_{s,p}^F$ are not positively definite. These terms can be decomposed using
\begin{eqnarray} \label{KinMC}
&&\int\limits_{z_1^m}^1 dz_1 \frac {1+z_1^2}{1-z_1} \biggl(K_s(z_1)\;\sigma _{1\gamma}(z_1)-\sigma _{1\gamma}\biggr)= \nonumber\\
&&\int\limits_{z_1^m}^{1-\frac{\Delta}{E}} dz_1 \frac {1+z_1^2}{1-z_1} K_s(z_1)\;\sigma _{1\gamma}(z_1) 
-\sigma _{1\gamma}\int\limits_{z_1^m}^{1-\frac{\Delta}{E}}  dz_1 \frac {1+z_1^2}{1-z_1}  \nonumber\\
&&+\int\limits_{1-\frac{\Delta}{E}}^1 dz_1 \frac {1+z_1^2}{1-z_1} \biggl(K_s(z_1)\;\sigma _{1\gamma}(z_1)-\sigma _{1\gamma}\biggr) \end{eqnarray}
and similarly for $\sigma_{p}^F$. The quantity $\Delta$ is defined as a minimal energy of the photon we want to generate (i.e., calorimeter resolution)
and $E=E_{beam}$.
The second integral  in (\ref{KinMC})  is calculated analytically. Third integral vanishes for $\Delta\rightarrow 0$, therefore it could be neglected (or kept and added to the contribution of the one-gamma contribution). The calculation  results in  
\begin{eqnarray} 
\sigma_s^F&=&\sigma_s(\Delta) + \delta_s(\Delta)\sigma_{1\gamma},
\nonumber\\
\sigma_p^F&=&\sigma_p(\Delta) + \delta_p(\Delta)\sigma_{1\gamma},
\end{eqnarray}
where $\sigma_{s,p}(\Delta)$ represent the first term in (\ref{KinMC}).

Combining all together we have for the cross section with the lowest order RC
\begin{equation}\label{CSforMC}
\sigma _{obs}=\Bigl( 1 + \frac{\alpha}{\pi}\bigl( \delta_{vac}+\delta(\Delta)\bigr)\Bigr) \sigma _{1\gamma} + \sigma_{s}(\Delta)+\sigma_p(\Delta)+\sigma_{add},
\end{equation}
where 
\begin{eqnarray}
\delta(\Delta)&=&\delta_s(\Delta)+\delta_p(\Delta)+\delta_{inf}+\delta_{fin}
\nonumber\\
&=&L\biggl(\frac{3}{2}+\log\biggl({4M^2\Delta^2\over SX}\biggr)\biggr).
\end{eqnarray}
Each contribution in (\ref{CSforMC}) is positively definite. The price for this representation is the dependence on $\Delta$. 

The event is generated for a kinematical point $x$, $Q^2$, $t$, and $\phi$  according to (\ref{CSforMC}), 
and the electron azimuthal angle ($\phi_e$) is simulated uniformly. 
Then 
 the probabilities of all three channels: nonradiated (i.e., no an additional radiated photon), radiated in $s$-peak, and radiated in $p$-peak are calculated 
 as:
\begin{eqnarray}
p_{nonrad}&=&\bigl( 1 + \frac{\alpha}{\pi}\bigl( \delta_{vac}+\delta(\Delta)\bigr)\bigr) \frac{\sigma _{1\gamma}}{\sigma_{obs}},
\nonumber\\
p_{s-peak}&=&\frac{\sigma_s(\Delta)}{\sigma_{obs}},
\nonumber\\
p_{p-peak}&=&\frac{\sigma_p(\Delta)}{\sigma_{obs}}.
\end{eqnarray}
The  scattering channel is generated according to these three probabilities. 
If the event with one photon in the final state
is chosen then no additional variables are needed to be simulated.
If the two-photon event in $s$-peak or $p$-peak is chosen,  
then the three kinematical variables of an additional photon are needed to be simulated. 
The photon energy is simulated through the variable $z_{1}$ or $z_{2}$ (for $s$- and $p$-peaks respectively) according to their distributions in integrand of $\sigma_s(\Delta)$ and $\sigma_p(\Delta)$. The photon angles are simulated in $s$- or $p$-peaks, i.e., the photon angles become equal to the angles of the initial or final lepton. 
Note, in (44) the components of $\sigma_{add}$ that correspond to $s$- and $p$-peaks are included in the definition of
$\sigma_s(\Delta)$ and $\sigma_p(\Delta)$. 

\section{Codes for Numerical Calculation of RC in a kinematical point and Monte Carlo generator}\label{codes}

The results presented in previous section allows us to create a code for numeric calculation of RC in a kinematical point (i.e., for specific $x$, $Q^2$, $t$, $\phi$, and beam energy) and respective Monte Carlo 
generator \footnote{The codes are available by request (email:ily@hep.by)}. 

The Fortran
code is called DVCSLL. Special keys allow to choose the part of the cross section (i.e., BH only, BH-DVCS interference, etc.), the approximation for hadronic part (exact for BH only or BMK), electron and proton  polarizations,  accuracies of integration, and the values of kinematical variables and the cut on missing mass. The Monte Carlo generator GenDVCSLL works as a slave system, i.e., generates one event for a kinematical point externally given. Additional parameter for the Monte Carlo generator is $\Delta$. 

Thus, DVCSLL is the code to calculate RC to BH process in leading approximation. The specific features of the approach and properties of the results include: i) the BH cross section of the lowest order in a shifted kinematical point is factorized in integrand, ii) no any assumptions about hadronic structure (except of choosing a specific form for nucleon form factors) are required, iii) cases of longitudinal and transverse target polarization are included, iv) higher order correction are included using a procedure of exponentiation (alternative approach in terms of electron structure functions was used in \cite{ByKuTo2008PRC, AkushevichIlyichev2012}), v) cut on 
missing energy
is implemented, and vi) both numeric and Monte Carlo integration methods are implemented. BMK approximation is used to describe
the hadronic structure for DVCS. Note that only leading log correction is implemented. 
The next-to-leading RC for the BH cross section of polarized particles is calculated in \cite{AISh2014}.

The leading log accuracy is the main uncertainty of theoretical calculation. Other theoretical uncertainties the researcher has to keep in mind include i) higher order corrections through exponentiation procedure (not so high effect is expected), ii) accuracy of numeric integration (largely under control), and iii)  approximations made when experimental cuts are implemented (could be tested) and finally resolved (using Monte Carlo generators). Besides there are physical contributions not taken into account yet, e.g.,  the pentagon (or 5-point) diagrams, i.e., the box diagram with a photon emission from the lepton line. Another type of uncertainty is the model dependence, such as the model for the nucleon formfactors (essential effect is not expected, but needed to be checked for each specific data analysis) and the BMK approximation for RC to DVCS. 

 The design of the Monte Carlo generator BHRadgen is as follows.
The input required by the generator
is: i) the four kinematical variables $x$, $Q^2$, $t$, and $\phi$, ii) the value of $\Delta$, iii) beam energy, and iv) the value of $V^2_{cut}$. The 
output is: 
i)  generated channel of scattering for an event, i.e., ``radiated'' (two photons in final state) or ``non-radiated'' (one photon in final state),  
ii) three additional kinematical variables (to describe an additional photon) generated for ``radiated'' event, and
iii) the cross section of RC for any event.
The cross sections and distributions over additional kinematical variables are calculated for the given kinematical point ($x$, $Q^2$, $t$, and $\phi$). Then any number of events are simulated using this information. If simulation of many events is required for a certain kinematical point, then the program  is efficient. However, the computation is not so fast if the point needs to be simulated for each event. 
 Approaches to accelerate generation of an event could include: i) a look-up table storing information about additional photon energies and angles in a kinematical region, ii) relaxation of requirements to the accuracy of Monte Carlo integration, and iii) using a numeric approach for integration and calculation of distribution over additional photonic variables.
 Collinear kinematics is used for simulation of photonic angles. 
 Instead, the distribution can be used from integrand over photonic angles.
 The calculation is based on the leading log approximation. 
 Next-to-leading corrections can be implemented using results for the RC calculation with the next-to-leading accuracy \cite{AISh2014}. 
In this case new analytical results for the distribution over additional photonic variables need to be obtained and implemented. Current code was obtained using the results integrated over two angles of an additional photon.
Exact formulae are implemented for the BH only. Contributions of DVCS are calculated in the BMK approximation.

\section{Numeric analysis}
\label{SectNumeric}
The experimental access to characteristics of the DVCS amplitudes is provided by the measurement of the beam spin asymmetry (\ref{eq11}). The observed asymmetry can be represented as
\begin{equation}
A=A_{1\gamma}{\delta_p \over \delta_u},
\end{equation}
where $\delta_{u,p}$ are RC factors for unpolarized (i.e., presented in the denominator of $A_{1\gamma}$) and polarized (i.e., presented in the nominator of $A_{1\gamma}$) parts of the cross section. The relative correction to asymmetry is defined as:
\begin{equation}
\delta_A={A-A_{1\gamma} \over A_{1\gamma}}.
\end{equation}

\begin{figure}\centering
\scalebox{0.5}{\includegraphics{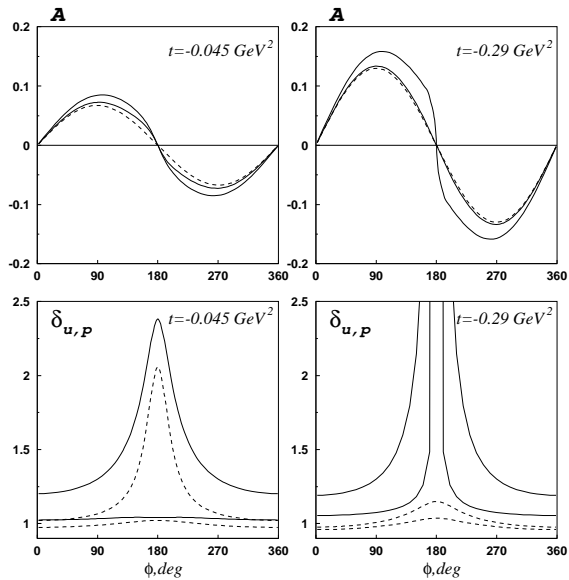}}
\caption{\label{asymdep}The $\phi$-dependence of the asymmetry (upper) and RC  factors (lower plots). Dashed curve at the upper plots gives the $\sigma_{1\gamma}$ and solid curved show the observed cross sections with 
$V_{cut}^2$=0.3 GeV$^2$ (the curve closer to dashed curve) and without cuts. Dashed and solid curves at bottom plots show $\delta_{u,p}$ with and without the cut, respectively. The curves with higher values corresponds to $\delta_p$, i.e., $\delta_p>\delta_u$.   Kinematical variables used for this example were $x$=0.1, $Q^2$=2GeV$^2$, and $E_{beam}$=11GeV.
  }
 \end{figure}
 
\begin{figure}\centering
\scalebox{0.5}{\includegraphics{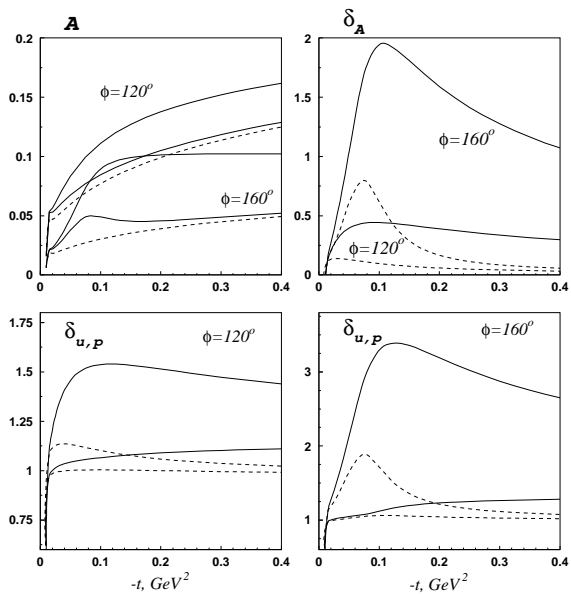}}
\caption{\label{tdep}The $t$-dependence of the asymmetry (upper left), RC to asymmetry (upper right) and RC  factors (lower plots). Dashed curve for $A$ gives the $\sigma_{1\gamma}$ and solid curved show the observed cross sections with $V_{cut}^2$=0.3 GeV$^2$ (the curve closer to dashed curve) and without cuts. Dashed and solid curves at the other three plots show $\delta_{A,u,p}$ with and without the cut, respectively.  Kinematical variables used for this example were $x$=0.1, $Q^2$=2GeV$^2$, and $E_{beam}$=11GeV.
  }
 \end{figure}
 The results for $\phi$- and $t$- dependencies are presented in Figure~\ref{asymdep} and Figure~\ref{tdep} respectively. RC is much higher in the region of 90$^o<\phi<$270$^o$ and small value of $-t$.  It is clear that the largest contribution to RC comes from collinear kinematics when $w_0$ or $u_0$ is minimal. Minimal values of these quantities can be achieved  in this region when $-t$ trends to its minimum values, $\cos \phi$ is negative and the absolute minimum is of order $m^2$ for 
$\phi$=180$^o$.

\section{Summary and Conclusion} 
\label{SectDiscussion}
The steps requiring for calculation of RC to BH and DVCS cross sections are: i) matrix element squared, ii) integration over loops and taking care on ultraviolet divergence (i.e., making the electron charge and mass renormalization), iii) phase space parameterization and integration over a part of kinematical variables of an additional photon: BH cross section is defined by four kinematical variables ($x$, $Q^2$, $t$ and $\phi$) and the cross section with two photon emitted is defined by seven kinematical variables, including the same 
four variables ($x$, $Q^2$, $t$ and $\phi$) and three additional variables (the two-photon invariant mass squared $ V^2$ and two angles of the photon pair), iv) 
extract and cancel the infrared divergence without making new assumptions, v) add a contribution of higher order corrections (calculated approximately), and vi) code the results to have, first, a program for RC calculation in a kinematical point ($x$, $Q^2$, $t$ and $\phi$) and Monte Carlo generator with inclusion of RC contributions. The calculation in the leading log approximation resulted in (\ref{iniint}). Adding the contribution of loops and cancelling the infrared divergence results in (\ref{eqeq30}). Finally the exponentiation of multiple soft photons results in the final formula in (\ref{exponentiation}). These formulae are valid for interference of BH and DVCS amplitudes and for pure DVCS contribution. The calculation of NLO correction to the BH cross section is much more complicated 
\cite{AISh2014}. The code created to calculate RC in a kinematical point based on these calculations is named DVCSLL. This code allows us to calculate RC to BH and DVCS process in leading approximation. Cases of longitudinal and transverse target polarization are included. Higher order corrections are included through exponentiation and potentially higher order corrections in terms of electron structure functions available for BH can be included for the interference of BH and DVCS as well. An opportunity to incorporate a cut on missing energy is implemented. Both numeric and Monte Carlo integration methods are implemented, integration over $\phi$ is implemented, and BMK approximation \cite{BKM2002} is used to describe hadronic structure for DVCS. The approach implemented in the Monte Carlo generator allowing for generation DVCS events. For each event the Monte Carlo generator selects between ``radiated'' (two photons in final state) or ``non-radiated'' (one photon in final state) and if ``radiated'' event is selected,   
three additional kinematical variables to describe an additional photon  are generated.
Numerical analysis of the RC to cross sections and asymmetries allowed us to conclude that the RC is under control and remaining uncertainties are due to model dependence and to the effects not taken into account yet (e.g., the Pentagon diagrams).

\appendix
\section{\label{sadd}} 
The analytical expression for the term $\sigma_{add}$ is:
\begin{eqnarray}\label{sigmaadd}
&&\sigma_{add}=
\frac{\alpha L}{2\pi}\Biggl[
\\
&&\qquad
\int\limits_{z_1^p}^{z_1^m}dz_1\frac{1+z_1^2}{1-z_1}
\frac{
{\hat \delta}_s\sin\theta^\prime_s+\tilde \delta_s\sin\tilde \theta^\prime_s
}{{\mathcal D}_{0s}^{1/2}}\left(\frac{x_s}{x} \right)^2\sigma_{1\gamma }(z_1)
\nonumber
\\&&
+
\int\limits_{z_2^p}^{z_2^m}dz_2\frac{1+z_2^2}{z_2(1-z_2)}
\frac{{\hat \delta}_p\sin\theta^\prime_p+\tilde\delta_p\sin\tilde \theta^\prime_p}{{\mathcal D}_{0p}^{1/2}}\left(\frac{x_p}{x} \right)^2\sigma_{1\gamma }(z_2)
\Biggr]
\nonumber
\end{eqnarray}
for $\cos \phi<0$, and $\sigma_{add}=0$ for $\cos \phi\ge0$. In (\ref{sigmaadd}), 
$\sigma_{1\gamma }(z_{1,2})$ are defined  by (\ref{sz1z2}), 
 $\sin\theta '$ is given in \cite{AkushevichIlyichev2012} by eq.~(38) as one of  solution of the eq.~(35),
$\sin{\tilde \theta}^\prime_{s,p}$ represent other solution:
\begin{eqnarray}
\sin{\tilde \theta}^\prime=-\frac{\cos\theta _z\sqrt{{\cal D}_0}+A\sin\theta_z\cos\phi}{\cos\theta_z^2+\sin^2\theta_z\cos^2\phi}.
\end{eqnarray}
The lowest limits of integration in (\ref{sigmaadd}) are defined as
\begin{eqnarray}
z^p_{1,2}&=&1-4\lambda_YV^2_+V^2_-[(V^2_--V^2_+)\sqrt{D_{s,p}}
\nonumber\\&&
+(V^2_++V^2_-)A_{2\; s,p}
+2V^2_+V^2_-(\lambda_Y+S_xS_p)]^{-1},
\nonumber\\&&
\end{eqnarray}
where:
\begin{eqnarray}
V^2_{\pm}&=&\frac{t S_x\pm\sqrt{\lambda_t}\sqrt{\lambda_Y}}{2M^2}-Q^2+t,
\nonumber\\
D_{s,p}&=&A_1^2V^2_+V^2_-+A_{2\; s,p}^2,
\nonumber\\
A_1&=&4M\cos(\phi)\sqrt{Q^2(S X-M^2Q^2)-m^2\lambda_Y},
\nonumber\\
A_{2\; s}&=&Q^2(S_p(S_x+2t)-\lambda_Y)-t(\lambda_Y+S_pS_x),
\nonumber\\
A_{2\; p}&=&Q^2(S_p(S_x+2t)+\lambda_Y)+t(\lambda_Y-S_pS_x).
\label{ad12}
\end{eqnarray}
The quantities ${\hat \delta}_{s,p}$ and $\tilde \delta_{s,p}$ 
are introduced to reflect experimental cuts on $V^2$ and therefore $z_{1,2}$. There are four cases for the cutting value of $V^2$, i.e., no cut and when  $V^2_{cut}$ is between $V_{1\;s,p}^2$, $V_{2\;s,p}^2$ or $V^2_{max}$ (Figure~\ref{zvv2rel}b). First, when no cut on missing mass square is used, ${\hat \delta}_{s,p}=\tilde \delta_{s,p}=1$. Second, if $V_{cut}^2\leq V_{1\;s,p}^2$, then ${\hat \delta}_{s,p}=\tilde \delta_{s,p}=0$. Third, if $V^2_{1\; s,p}<V^2_{cut}<V^2_{2\; s,p}$, then ${\hat \delta}_{s,p}=\theta(z_{1,2}-z_{1,2}^c)$ and $\tilde \delta_{s,p}=0$. Fourth, $V^2_{2\; s,p}<V^2_{cut}<V^2_{max}$ then ${\hat \delta}_{s,p}=1$ and $\tilde \delta_{s,p}=\theta(z_{1,2}^c-z_{1,2})$. Formally ${\hat \delta}_{s,p}$ and $\tilde \delta_{s,p}$ can be presented using a combined formula aggregating all four cases:
\begin{eqnarray}
{\hat \delta}_{s,p}&=&\theta(V^2_{cut}-V^2_{1\; s,p})\theta(V^2_{2\;s,p}-V^2_{cut})\theta(z_{1,2}-z_{1,2}^c)
\nonumber\\ &&
+\theta(V^2_{cut}-V^2_{2\; s,p}),
\nonumber\\
\tilde \delta_{s,p}&=&\theta(V^2_{cut}-V^2_{2\; s,p})\theta(z_{1,2}^c-z_{1,2}).
\label{delt}
\end{eqnarray}
The restrictions on $z_{1,2}$ read:
\begin{eqnarray}
z_{1,2}^c&=&1-2\lambda_ YV_{cut}^2[A_1\sqrt{(V^2_+-V^2_{cut})(V^2_{cut}-V^2_-)}
\nonumber\\&&
+A_{2\; s,p}+V_{cut}^2(S_pS_x+\lambda_q)]^{-1}
\label{z12c}
\end{eqnarray}
and the other quantities used in previous expressions are:
\begin{eqnarray}
V^2_{1\; s,p}&=&\frac{V^2_+D_{s,p}}{A_1^2V^2_++A_{2\; s,p}^2}<V^2_{max}, \nonumber \\
V^2_{2\; s,p}&=&\frac{2V^2_+V^2_-\sqrt{D_{s,p}}}
{(V^2_++V^2_-)\sqrt{D_{s,p}}-A_{2\; s,p}(V^2_+-V^2_-)   }.
\end{eqnarray}

\vspace{1cm}
\noindent {\bf Acknowledgments}. The authors are grateful to Harut
Avakian and Volker Burkert for interesting discussions
and comments. This work was supported by DOE contract
No. DE- AC05-06OR23177, under which Jefferson
Science Associates, LLC operates Jefferson Lab.

\bibliography{dvcsll}{}

\end{document}